%% file: GBb.tex
\newcommand*{\ATLASLATEXPATH}{latex/}
\documentclass[cernpreprint,texlive=2011,txfonts,UKenglish]{latex/atlasdoc}
\usepackage{latex/atlaspackage}
\usepackage{latex/atlasphysics}

\usepackage{\ATLASLATEXPATH atlaspackage}
\usepackage{multirow}
\usepackage{float}
\usepackage{\ATLASLATEXPATH atlasbiblatex}

\usepackage{\ATLASLATEXPATH atlascontribute}

\usepackage{\ATLASLATEXPATH atlasphysics}

\usepackage{color}
\usepackage{xcolor,colortbl}

\graphicspath{{logos/}{figures/}}

\input{GBb-metadata}
\hypersetup{pdftitle={ATLAS draft},pdfauthor={The ATLAS Collaboration}}

\newcommand{\Glu}{$G^*$\xspace}
\newcommand{\cellgreen}{\cellcolor{green!60}}
\newcommand{\cellyellow}{\cellcolor{yellow!70}}
\newcommand{\cellred}   {\cellcolor{red!65}}
\newcommand{\largeR}    {large-\ensuremath{R}\xspace}
\newcommand{\LargeR}    {Large-\ensuremath{R}\xspace}
\newcommand{\smallR}    {small-\ensuremath{R}\xspace}
\newcommand{\SmallR}    {Small-\ensuremath{R}\xspace}
\newcommand{\antikt}            {anti-$k_{t}$\xspace}

\newcommand{\kt}                {\ensuremath{k_{t}}\xspace}

\begin{document}

\maketitle



\input{Introduction.tex}
\input{atlas-detector}
\input{simulation}
\input{object}

\input{selection}
\input{background}
\input{systematics}
\input{Results}


\input{Conclusion}

\clearpage
\section*{Acknowledgements}

\input{acknowledgements/Acknowledgements}

%

\clearpage


\printbibliography
\clearpage
\input{atlas_authlist}

%

\end{document}

%% file: GBb-metadata.tex
\AtlasTitle{Search for single production of a vector-like quark via a heavy gluon in the $4b$ final state with the ATLAS detector in $pp$ collisions at $\sqrt{s} = 8$~TeV}

\author{The ATLAS Collaboration}

\date{\today}

\AtlasRefCode{EXOT-2014-19}

\AtlasNote{EXOT-2014-19}

\PreprintIdNumber{CERN-PH-EP-2015-321}

\AtlasJournal{Phys.\ Lett.\ B.}

\AtlasAbstract{%
A search is performed  for the process $pp \to G^* \to B_H\bar b/\bar B_H b \to H b \bar b \to b\bar bb\bar b$, predicted in composite Higgs scenarios, where $G^*$ is a heavy colour octet vector resonance and $B_H$ a vector-like quark of charge $-1/3$. The data were  obtained from $pp$ collisions at a centre-of-mass energy of $8$~TeV corresponding to an integrated luminosity of $19.5\text{ fb}^{-1}$, recorded by the ATLAS detector at the LHC. The largest background, multijet production, is estimated using a data-driven method. No significant excess of events with respect to Standard Model predictions is observed, and upper limits on the production cross section times branching ratio are set.  Comparisons to the predictions from a specific benchmark model are made, resulting in lower mass limits in the two-dimensional mass plane of $m_{G^*}$ $\text{vs.}$ $m_{B_H}$.}

%% file: Introduction.tex
\section{Introduction}
\label{sec:introduction}
Composite Higgs~\cite{Dugan:1984hq,Agashe:2004rs,Contino:2003ve,Contino:2006qr}  models interpret the Higgs boson discovered at the Large Hadron Collider (LHC)~\cite{LHC:JINST} as a pseudo-Goldstone boson resulting from spontaneous symmetry breaking in a new strongly coupled sector, thus addressing the naturalness problem, the extreme fine tuning required in the Standard Model (SM) to cancel quadratically divergent radiative corrections to the Higgs boson mass.  A generic prediction of these models is the existence of massive vector-like quarks (VLQ). These VLQs are expected to mix mainly with the third family of quarks of the SM~\cite{Kaplan:1991dc,Contino:2006nn,Contino:2011np}, leading to partial compositeness. Colour octet resonances (massive gluons) also occur naturally in these models~\cite{Kaplan:1991dc,Contino:2006nn,Bini:2011zb,Chala:2014mma}. 

\begin{figure}[htb]
	\begin{center}
		\includegraphics[width=3.04in]{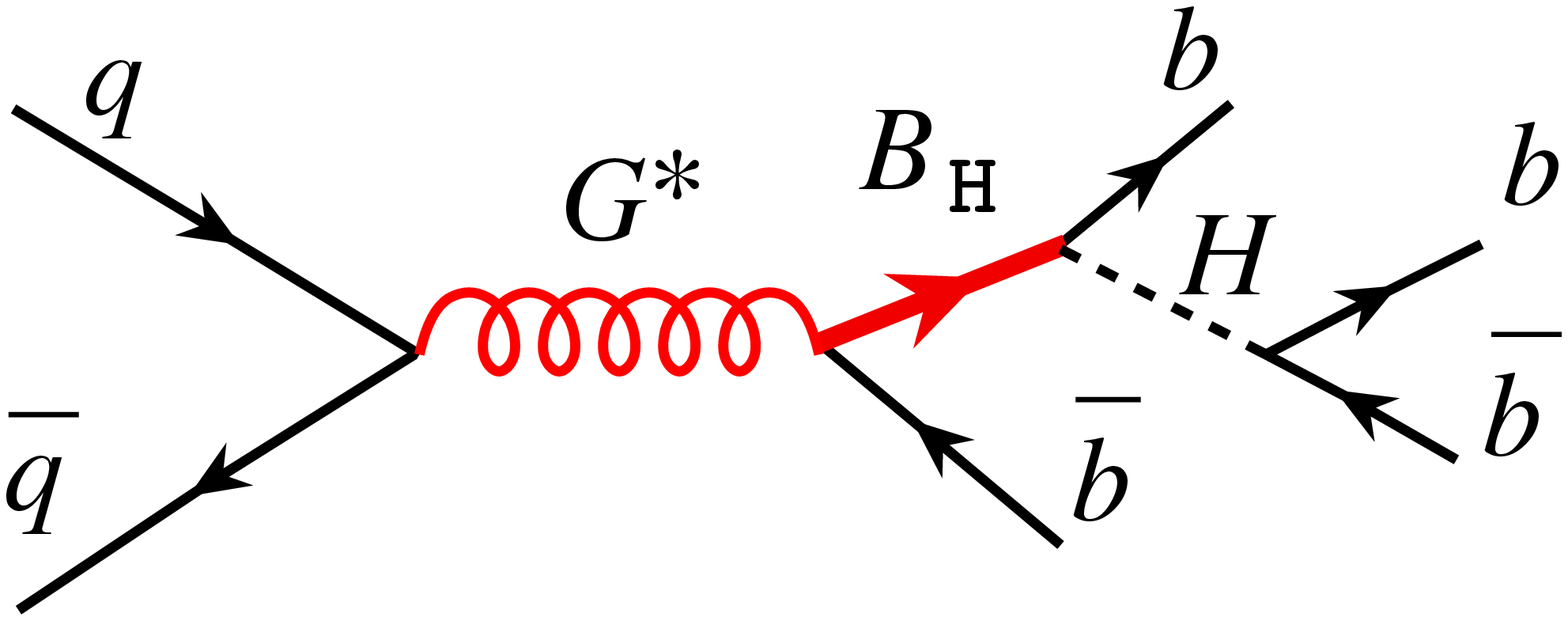} 
	\end{center}
	\caption{Feynman diagram of the signal process $q \bar q \to G^* \to B_H\bar b \to H b \bar b \to b\bar bb\bar b$.}
	\label{fig:GBb}
\end{figure}

Searches for vector-like quarks in the ATLAS and CMS experiments, in both the pair and single production processes ~\cite{ATLAS:2012qe,EXOT-2013-17,Aad:2015kqa,Aad:2015mba,Aad:2015gdg,Aad:2015tba,Aad:2015voa,CMS-B2G-12-015,Chatrchyan:2013wfa,Khachatryan:2015gza,Khachatryan:2015oba,Khachatryan:2015axa}, 
constrain their mass to be above $700$--$900$~GeV. This analysis is a search for single production of a vector-like quark $B_H$ of charge $-1/3$ via the $s$-channel exchange of  a heavy colour octet vector resonance \Glu, using data recorded by the ATLAS detector at the LHC. The search is performed for the process of $Hb\bar b$ production through  $pp \to G^* \to B_H\bar b/\bar B_H b \to H b\bar b \to b\bar bb\bar b$ (see Fig.~\ref{fig:GBb}),\footnote{Charge conjugate states are implied in the following text.} based on Ref.~\cite{Chala:2013ega} and using the benchmark model of Ref.~\cite{Bini:2011zb}. 
This simplified minimal composite Higgs model has a composite sector with a global $\text{SU}(3)_c \times \text{SU}(2)_\text{L} \times \text{SU}(2)_\text{R} \times \text{U}(1)_Y$ symmetry and an elementary sector which contains the SM particles but not the Higgs boson. Physical states of the composite sector include the heavy gluon $G^*$, a composite Higgs boson and heavy vector-like quarks of charge $5/3$, $2/3$, $-1/3$ and $-4/3$. Among these heavy quarks, there is one singlet of charge $2/3$ which mixes with the right-handed top quark of the SM with an angle $\theta_{t_{\rm R}}$, and similarly one singlet of charge $-1/3$ which mixes with the right-handed bottom quark of the SM with an angle $\theta_{b_{\rm R}}$. After mixing between the gluons from the elementary and composite sectors by an angle $\theta_s$,  the physical state of the heavy gluon has a coupling  $g_c \cos\theta_s $ to composite states, where $g_c = g_s/\sin\theta_s$ and $g_s$ is the coupling of the SM gluon. The other parameters of the model are the composite fermion masses,  assumed to be universal, the heavy gluon mass $m_{G^*}$ and two Yukawa couplings $Y_T$ and $Y_B$. In a large part of the parameter space, the lightest of the new heavy quarks is  $B_H$, of charge $-1/3$, and in this model it decays exclusively to $H b$. In Ref.~\cite{Chala:2013ega}, the condition $m_{B_H} = m_{G^*}/2$ is applied, with the result that pair production of the heavy partners is kinematically forbidden and the width of $G^*$ is consequently not too large. In the search presented here, the phase space is extended to $m_{B_H} \ge m_{G^*}/2$. When  $m_{B_H} < m_{G^*}/2$, present results on pair production of vector-like quarks can be recast in a model with a massive colour octet~\cite{Araque:2015cna}.

For high masses of the \Glu and $B_H$ resonances, the Higgs boson is highly boosted and the decay products are reconstructed in a single large-radius (\largeR) jet in the detector,  whereas for lower masses the four $b$-quarks are reconstructed as separate small-radius jets. The analysis uses two sets of selection criteria to target these two cases.

%% file: atlas-detector.tex
\section{The ATLAS detector}
\label{sec:detector}
The ATLAS detector,\footnote{ATLAS experiment uses a right-handed coordinate system with its origin at the nominal interaction point in the centre of the detector and the $z$-axis along the beam pipe. The $x$-axis points from the IP to the centre of the LHC ring, and the $y$-axis points upward. Cylindrical coordinates $(r,\phi)$ are used in the transverse plane, $\phi$ being the azimuthal angle around the $z$-axis. The pseudorapidity is defined in terms of the polar angle $\theta$ as $\eta=-\ln\tan(\theta/2)$. The distance in $\eta$--$\phi$ space is referred to as $\Delta R = \sqrt{\left(\Delta\eta\right)^2 + \left(\Delta\phi\right)^2}$.} located at the LHC, is described in detail in Ref.~\cite{PERF-2007-01}. It covers nearly the full solid angle around the collision point. The inner detector is surrounded by a solenoid that produces a 2~T axial magnetic field. The tracks of charged particles are reconstructed with a high-granularity silicon pixel and microstrip detector for $|\eta|<2.5$. A straw-tube transition radiation detector extends the tracking to larger radii and provides electron/pion discrimination. The electromagnetic calorimeter consists of a barrel and end-cap lead/liquid-argon (LAr) sections with an accordion geometry covering $|\eta|<3.2$, preceded by a thin presampler, covering $|\eta|<1.8$, which allows corrections for fluctuations in upstream energy losses. A copper/LAr electromagnetic calorimeter covers the very forward angles. 
Hadronic calorimetry is installed in the barrel region, $|\eta|<1.7$, using steel as the absorber and scintillator tiles as the active material. In the endcaps, copper/LAr calorimeters cover $1.5<|\eta|<3.2$ followed by a forward calorimeter based on tungsten absorbers in LAr as sensitive medium, up to $|\eta| = 4.9$. Surrounding the hadronic calorimeters are large toroidal magnets whose magnetic fields deflect the trajectories of charged particles exiting the barrel and end-cap calorimeters. The muon spectrometer uses monitored drift tubes for tracking in $|\eta|<2.7$ with cathode strip chambers in the innermost station for $|\eta|>2.0$. A dedicated muon trigger is provided by resistive plate chambers in the barrel and thin-gap chambers in the end-cap, covering $|\eta|<2.4$.

A three-level trigger system, consisting of a hardware Level-1 trigger and two software-based trigger levels reduce the event rate to be recorded to less than about 400~Hz.

%% file: simulation.tex
\section{Data and simulation}
\label{sec:simulation}
Data used in this analysis correspond to an integrated luminosity of $19.5\text{ fb}^{-1}$ of $pp$ collisions collected at the LHC at a centre-of-mass energy of \sqs = 8 TeV, with all the essential elements of the ATLAS detector fully operational and stable.

Simulated signal and background samples are produced by Monte Carlo (MC)  event generators and passed through a \GEANT 4~\cite{Agostinelli:2002hh} simulation of the ATLAS detector~\cite{SOFT-2010-01}. Additional events from the same and neighbouring bunch crossings (pile-up) are included by adding simulated diffractive and non-diffractive $pp$ collisions to hard-scattering events. The pile-up rate is reweighted in accordance with the luminosity profile of the recorded data. All simulated events are then reconstructed using the same reconstruction software as the data.

Signal samples based on the model discussed in Ref.~\cite{Chala:2013ega} are generated with \textsc{MadGraph}5\_aMC@NLO~\cite{Alwall:2014hca}, using CTEQ6L1~\cite{Nadolsky:2008zw} parton distribution functions (PDFs), in the mass region $m_{G^*}/2 \le m_{B_H} < m_{G^*}$, with $1~\text{TeV} < m_{G^*} < 3$~TeV, in steps of 250 GeV in $m_{G^*}$ and in steps of 125~GeV in $m_{B_H}$. The Higgs boson mass is set to 126~GeV and its branching ratio $\text{BR}\left(H\to b\bar b\right)$ to 56.1\%~\cite{Heinemeyer:2013tqa}. The parameters of the model are set as in Ref.~\cite{Chala:2013ega}: $g_c =3$, $Y_T=Y_B=3$, $\sin\theta_{t_{\rm R}}=\sin\theta_{b_{\rm R}}= 0.6$.

The event selection requires at least two $b$-jets in the final state. Multijet events from strong interactions have a large cross section and are the dominant background. Due to the large number of events required to simulate this background and the difficulty of modelling it accurately, it is evaluated using a data-driven method, as described in Section 6. Other background contributions include top-pair and single-top-quark production, generated with {\sc Powheg-Box} ~\cite{Nason:2004rx,Frixione:2007vw,Alioli:2010xd} interfaced to \PYTHIA~\cite{Sjostrand:2006za} using CT10 PDFs~\cite{Lai:2010vv}. The $t\bar t$ sample is normalised to the theoretical calculation performed at next-to-next-to-leading order (NNLO) including resummation of next-to-next-to-leading logarithmic (NNLL) soft gluon terms with \textsc{Top}++2.0~\cite{Czakon:2013goa,Czakon:2011xx},  giving an inclusive cross section of $253^{+13}_{-15}$ pb~\cite{ttbarxsec}. Samples of $t\bar t + Z$ and $t\bar t +H$ events are generated with \PYTHIA and CTEQ6L1 PDFs. The \SHERPA~\cite{Gleisberg:2008ta} generator, with CT10 PDFs, is used to simulate $W/Z + \text{jets}$ samples with leptonic decay of the vector bosons. \SHERPA is also used to generate $Z+\text{jets}$ events, with $Z \to b \bar b$, where the extra jets are produced inclusively. Contributions from diboson backgrounds---$WW$, $WZ$ and $ZZ$---are estimated to be negligible.

%% file: object.tex
\section{Object reconstruction}
\label{sec:object}
The final state consists of four jets from $b$-quarks ($b$-jets), two of which come from the Higgs boson decay. If the Higgs boson is sufficiently boosted, having a transverse momentum $\pt \gtrsim300 $ GeV, the two $b$-jets may be merged into a single jet with a large radius parameter (\largeR jet) and therefore two different jet definitions are used.
 
Jets with smaller radius parameter, or \smallR jets, are reconstructed from calibrated calorimeter energy clusters~\cite{ATL-LARG-PUB-2008-002,Cojocaru:2004jk} using the \antikt algorithm~\cite{Cacciari:2008gp} with a distance parameter $R=0.4$. The high \pt threshold used in the event selection ensures that the contamination of jets from pile-up is small. To ensure high-quality reconstruction of central jets while rejecting most jets not coming from hard-scattering events, criteria as described in Ref.~\cite{ATLAS-CONF-2010-038} are applied. Jets are corrected for pile-up by a jet-area subtraction method and calibrated by a jet energy scale factor~\cite{PERF-2011-03}. They are required to have $\pt > 50$~GeV and $\abseta < 2.5$.

\SmallR jets are identified as containing a $b$-hadron ($b$-tagged) by a multivariate algorithm~\cite{ATLAS-2015ydr}.  This algorithm was configured to give a $b$-tagging efficiency of 70\% in simulated $t\bar t$ events, with a mistag probability of about 1\% for gluon and light-quark jets and of about 20\% for $c$-quark-initiated jets. The $b$-tagging efficiency in simulated events is corrected to account for differences observed between data and simulation.

\LargeR jets are reconstructed using the \antikt algorithm with $R = 1.0$. Jet trimming~\cite{Krohn:2009th,PERF-2012-02} is applied to reduce the contamination from pile-up and underlying-event activity: subjets are formed using the \kt algorithm~\cite{Catani:1993hr} with $R = 0.3$ and subjets with $\pt(\text{subjet})/\pt(\text{jet}) < 5\%$ are removed.

Leptons are vetoed in this analysis to reduce background involving leptonically decaying vector bosons. Electron candidates with $\pt > 7$~GeV are identified in the range $\abseta < 2.47$ from energy clusters in the electromagnetic calorimeter, matched to a track in the inner detector. Requirements of `medium' quality, as defined in Ref.~\cite{PERF-2010-04}, are applied together with two isolation criteria: the scalar sum of the transverse momentum (energy) within a radius $\Delta R = 0.2$ around the electron candidate has to be less than 15\% (14\%) of the electron \pt $\left(E_\text{T}\right)$. Muons with $\pt > 7$~GeV and $\abseta < 2.4$ are reconstructed from matched tracks in the muon spectrometer and the inner detector. Quality criteria are applied, as described in Ref.~\cite{PERF-2014-05}, and an isolation requirement is applied: the scalar sum of the transverse momentum of tracks within a radius $\Delta R = 0.2$ around the muon candidate has to be less than 10\% of the muon \pt.

%% file: selection.tex
\section{Event selection}
\label{sec:selection}

Because of the very high hadronic background at the LHC, it is not possible to have adequate Monte Carlo statistics for multijet events. The uncertainties in the quality of simulation of $b$-jets at high-\pt can also be large. For these reasons, for each mass pair $\left(m_{G^*},m_{B_H}\right)$ being tested, a data-driven technique was used to evaluate the expected background, as described in Section~\ref{sec:background}. The technique requires that we define control regions orthogonal to the signal regions. A blind analysis is performed, in which the background is first evaluated without initial knowledge of the data in the signal regions. In order to test the large number of mass pair hypotheses, all signal region cuts are applied except the Higgs mass window which is blinded when evaluating the background in the signal regions.

\subsection{Event preselection}
Events in the signal region are first preselected according to the following criteria (see end of Section~\ref{sub:merged} for the signal region definition).
\begin{itemize}
	\item They must satisfy a combination of six triggers requiring multiple jets and $b$-jets for various \pt thresholds, where $b$-jets are identified by a dedicated online $b$-tagging algorithm. This combination of triggers is $>99\%$ efficient for signal events passing the offline selection, across the $B_H$ and $G^*$ mass ranges considered in this analysis.
	\item They are vetoed if they contain reconstructed isolated leptons $\left(e \text{ or } \mu\right)$ in order to reduce the contribution from $W/Z+\text{jets}$ and $t\bar t$ backgrounds.
	\item At least three \smallR  $b$-tagged jets must be present in the signal region.
	\item The invariant mass of the system composed of all selected $R = 0.4$ jets is required to be greater than 600~GeV.
\end{itemize}
Two event topologies are considered for the signal, depending on the boost of the Higgs boson. Highly boosted Higgs bosons are reconstructed using \largeR jets as described in Section~\ref{sec:object} and this topology corresponds to the merged scenario (see Section~\ref{sub:merged}). If no \largeR jet is found, an attempt is made to reconstruct the Higgs boson from two \smallR jets (see Section~\ref{sub:resolved}). The acceptance times reconstruction efficiency for the combined yields of the two topologies varies from 5\% to 20\% depending on the masses of the \Glu and $B_H$.

\subsection {Merged selection}
\label{sub:merged}
The signal region for the merged case consists of the following requirements.
\begin{itemize}
	\item A \largeR jet must be present with \pt > 300 GeV and \abseta < 2.0 and mass  in the range $\left[90, 140\right]$~GeV. The
mass window was optimised based on the signal sensitivity. If more than one such \largeR jet is present, the Higgs candidate is chosen to be the one with mass closest to 126~GeV.  At least one $b$-tagged jet must be matched to it within a distance $\Delta R = 1.0$. 
	\item There must be at least two additional $b$-tagged jets separated from the Higgs boson candidate, $\Delta R\left(H, j\right) > 1.4$. The two with the highest \pt are used to reconstruct the \Glu and $B_H$ candidates. 
\end{itemize}

Once the Higgs boson candidate has been identified as above, there remains an ambiguity in assigning the other jets to the vector-like quark $B_H$. The four-momentum of the $B_H$ candidate is reconstructed as the four-momentum sum of the Higgs boson candidate and either the next-to-leading-\pt (category 1) or the leading-\pt (category 2) $b$-jet away from it, depending on the assumed mass difference between \Glu and $B_H$. For large $G^*$--$B_H$ mass difference, the $B_H$ and $b$-quark from \Glu splitting have high momentum and therefore the jet from the subsequent $B_H$ decay is likely to be the next-to-leading jet. For a small mass difference the opposite is true since in this latter case the $B_H$ decay products are more boosted than the \Glu splitting products. For each $\left(m_{G^*},m_{B_H}\right)$ pair, the category which has the higher probability that the correct pairing is formed is chosen, based on the simulated signal events.  
Finally, the \Glu four-momentum is reconstructed as the four-momentum sum of the Higgs boson jet and the two leading-\pt $b$-jets not matched to the Higgs boson candidate.

Different signal regions are defined for the different $\left(m_{G^*},m_{B_H}\right)$ mass pair hypotheses. They are characterised by the choice of category defined above as well as by lower cuts on the reconstructed masses of \Glu and $B_H$ candidates. Five inclusive signal regions were defined, with the minimum mass of the \Glu candidate ranging from 0.8 to 1.8~TeV and of the $B_H$ candidate from 0.5 to 1~TeV; these are shown in Table~\ref{tab:sr_cuts}. No upper cut on the resonance masses was set since the multijet background distribution falls rapidly and the resonance widths become larger for high masses. For each mass pair considered, the signal region that gives the maximum signal sensitivity, the ratio of the expected number of signal events to the square root of the number of background events, is chosen.

\begin{table}[H]
	\begin{center}{\small
	\scalebox{0.88}{
	\begin{tabular}{c|c|c|c|c|c}
		\hline
		\hline
		& \multicolumn{2}{c|}{\textbf{Category 1}} & \multicolumn{3}{c}{\textbf{Category 2}}\\
		\cline{2-6}
                	& \textbf{SR1} & \textbf{SR2} & \textbf{SR3} & \textbf{SR4} & \textbf{SR5}\\
                	\hline
                	Lower cut on reconstructed $m_{G^*}$ and $m_{B_H}$ [TeV] & (1.0, 0.5) & (1.3, 0.5) & (0.8, 0.5) & (1.5, 0.5) & (1.8, 1.0)\\
                	\hline
	\end{tabular}}
	\caption{Signal region definitions:  category 1 (2) refers to the case where the next-to-leading-\pt (leading-\pt) jet not associated with the Higgs boson is assumed to be from the $B_H$ decay.} 
	\label{tab:sr_cuts}}
	\end{center}
\end{table}

\subsection {Resolved selection}
\label{sub:resolved}
Events in the resolved signal region are required to satisfy the following criteria.

\begin{itemize}
	\item In order to be able to later combine the results with the merged channel, events are required to fail the merged selection criteria.
	\item Events are required to have exactly four \smallR jets with $\pt > 50$~GeV and $\abseta < 2.5$, with at least three of these jets being $b$-tagged. The Higgs boson candidate is reconstructed using the two jets with invariant mass nearest to 126~GeV. The invariant mass is required to be in the interval $[90, 140]$~GeV and the transverse momentum of the dijet system $\pt(jj) > 200$~GeV.
\end{itemize}
The four-momentum of the $B_H$ candidate is reconstructed from the four-momentum sum of the Higgs candidate and either the leading or the next-to-leading-\pt jet away from the Higgs boson jets, depending on the $G^*$--$B_H$ mass splitting. As in the merged case, for each pair of masses considered the category is chosen to be the one with the lower mis-assignment rate of jets, based on samples of simulated signal events. Inclusive signal regions are defined by lower minimum mass values identical to the merged case, and shown in Table~\ref{tab:sr_cuts}. Each mass pair is assigned to the same SR for the merged and resolved analysis. The four-momentum of the $G^*$ candidate is reconstructed from the four-momentum sum of the four jets in the event. 

%% file: background.tex
\section{Modelling of the multijet background}
\label{sec:background}

The `ABCD' data-driven method is used to estimate the multijet background. For each of the ten signal regions, three control regions orthogonal to the signal region are defined: region~B has all the signal region selection criteria mentioned in Section ~\ref{sec:selection} applied, including the lepton vetoes and lower cuts on the masses of $B_H$ and $G^*$ candidates, but the Higgs boson candidate mass is required to be outside the interval $\left[90,140\right]$~GeV; region~C has all the signal region selection requirements, but requires exactly two jets to be $b$-tagged; and region~D has the Higgs boson candidate outside the Higgs boson mass window and exactly two $b$-tagged jets. In regions~C and D, only one of the two jets not associated with the Higgs boson candidate is $b$-tagged. The number of multijet (MJ) events expected in the signal region (SR) is then evaluated according to
\begin{equation}
	N^{\text{MJ}}_{\text{SR}}  = N_{\text{B}}/N_{\text{D}} \times N_{\text{C}},
	\label{eq:ABCD}
\end{equation}
where $N_{\text{X}}$ is the number of events in region X, after having removed the top-quark, diboson and other electroweak background contributions as determined from MC simulations. 

This estimate assumes that no bias results from the choice of control regions. To evaluate and potentially correct for the effect of any biases, a re-weighting is performed on two kinematic distributions, the leading-jet $\pt$ and the $\Delta R$  between the reconstructed Higgs boson candidate and the leading jet not associated with it. Control regions C and D (B and D) are re-weighted, using a method similar to Ref.~\cite{EXOT-2014-11}, to have the same shape as in control region B (C) with weights obtained from $N_{\text{B}}/N_{\text{D}}$ $\left(N_{\text{B}}/N_{\text{D}}\right)$ per bin. The effect of this re-weighting is found to be negligible and therefore no correction is applied.

A validation region is defined as the 15 GeV sideband regions outside the Higgs boson candidate mass window, i.e. 75-90 GeV and 140-155 GeV, for each signal region. The contribution from multijet background is estimated as above, but with the control regions B and D excluding these validation regions and region C now being the two sidebands. It is then compared to the number of observed data events, after adding back the simulation-based background, in these regions. Table~\ref{tab:vr} shows that the expected and observed numbers of events agree well in the validation regions for the merged- and resolved-channel signal regions.

\begin{table}[H]
        \begin{center}
        \begin{tabular}{c|c|c|c|c|c}
                \hline
                \hline
                \multicolumn{6}{c}{\textbf{Merged}}\\
                \hline
                				& \textbf{VR1}	& \textbf{VR2}	& \textbf{VR3}		& \textbf{VR4}	& \textbf{VR5}\\
                \hline
                Expected		& $563\pm16$	& $213\pm10$	& $1680\pm29$	& $135\pm8$	& $45\pm4$\\
                Observed		& $558$		& $184$		& $1666$			& $137$		& $35$\\
                \hline
                \hline
                \multicolumn{6}{c}{\textbf{Resolved}}\\
                \hline
                				& \textbf{VR1}		& \textbf{VR2}	& \textbf{VR3}		& \textbf{VR4}	& \textbf{VR5}\\
                \hline
                Expected		& $1065\pm21$	& $337\pm11$	& $3758\pm50$	& $242\pm10$	& $63\pm5$\\
                Observed		& $1073$			& $324$		& $3906$			& $238$		& $56$\\
                \hline
	\end{tabular}
         \caption{Expected and observed numbers of events in the validation regions (VR) associated to their respected signal regions for the merged and resolved channels. Only the statistical error is shown.}
        	\label{tab:vr}
        	\end{center}
\end{table}

%% file: systematics.tex
\section{Systematic uncertainties}
\label{sec:systematics}

Systematic uncertainties from several sources affect the expected numbers of background and signal events. Table~\ref{tab:syst_tot_resolved_ortho} shows the estimated size of the different components.

The statistical uncertainty in the data control regions used for the estimation of the multijet background is considered as part of the statistical error.

There is an uncertainty in the number of background events due to the difference between the observed and estimated numbers of events in each of the validation regions. In each validation region, if the observed number of events is compatible with the estimated number within one standard deviation (calculated as the sum in quadrature of the relative statistical errors of the two), this standard deviation is considered to be the background estimation uncertainty. 
Otherwise, the background uncertainty is considered to be the fractional difference between the observed and estimated numbers of events. This is the largest uncertainty, ranging from 5\% in SR1 to 27\% in SR5 for the merged case, and from 3.5\% in SR1 to 16\% in SR5 for the resolved case.

The $t\bar t$ contribution dominates the simulation-based background. The theoretical uncertainty on its cross section is taken to be 6\%, as discussed in Section~\ref{sec:simulation}.

Uncertainties due to the calibration and modelling of the detector affecting the simulation-based background estimates in the control and signal regions are principally due to the jet energy scale (JES) and jet energy resolution (JER). JES uncertainties for \smallR jets include contributions from detector reconstruction and from different physics modelling and evaluation methods~\cite{PERF-2012-01}. Uncertainties leading to a higher (lower) yield than the nominal value are added in quadrature to the total JES up (down) uncertainty. To evaluate the impact of JER for \smallR jets, energies of simulated jets are smeared to be consistent with the JER measured in data. The JER systematic uncertainty is the difference between the nominal and smeared values.

JES uncertainties for \largeR jets in the central region are evaluated as described in Ref.~\cite{PERF-2012-02}. The jet mass scale (JMS) uncertainty is 4--5\% for $\pt\lesssim700$~GeV and increases linearly with \pt to about 8\% in the range $900\lesssim\pt\lesssim1000$~GeV.


The total uncertainty in the measured $b$-tagging efficiency was evaluated in Ref.~\cite{ATLAS-CONF-2014-004} and is \pT and $\eta$ dependent. For high-\pt jets, the systematic uncertainty is derived from simulation. It is estimated here for the simulation-based backgrounds, accounting for the statistical uncertainty, the error on the generator-dependent scale factors, the track momentum scale, resolution and efficiency systematic uncertainties, and the extrapolation uncertainties for light jets. It is at or below the percent level and always dominated by the background estimation.

The predicted signal is not confined to the signal region: it could also constitute a fraction of the observed data in the control regions. The effect of this potential contamination on the statistical procedure is described in Section~\ref{sec:result}.

Systematic uncertainties due to detector effects also affect the VLQ signal yields. They are dominated by the $b$-tagging uncertainties, ranging from 16\% to 40\% depending on \pt, while other sources of systematic uncertainties listed above are below 5\%. Theoretical uncertainties in the signal cross section due to the choice of PDFs are estimated from CTEQ6.6 with its 22 eigenvector sets~\cite{Nadolsky:2008zw}.

\begin{table}[H]
        \begin{center}{\small
        \scalebox{0.9}{%
        \begin{tabular}{c|c|c|c|c|c}
             \hline
             \hline
             \multicolumn{6}{c}{\textbf{Merged}}\\
             \hline
             \textbf{Systematic uncertainty} & \textbf{SR1} & \textbf{SR2} & \textbf{SR3} & \textbf{SR4} & \textbf{SR5}\\
             \hline
             Background estimation	& 5\%	& 15\%	& 2.8\%	& 10\%	& 27\% \\ 
             $t\bar t$	 cross section	& $+1.0$\% $-1.1$\%	& $+0.8$\% $-0.9$\%	& $+1.2$\% $-1.4$\%	& $+0.8$\% $-0.9$\%	& $+0.6$\% $-0.7$\% \\
             JER \smallR		& $+0.29$\%	& $+0.15$\%	& $+0.01$\%	& $-0.32$\%	& $+0.20$\% \\
             JES \smallR		& $+0.9$\% $-0.8$\%	& $+1.6$\% $-0.7$\%	& $+1.0$\% $-1.0$\%	& $+0.9$\% $-1.0$\%	& $+1.5$\% $-1.0$\% \\
             JES/JMS \largeR	& $+0.31$\% $-1.5$\%	& $+1.3$\% $-1.5$\%	& $+0.13$\% $-1.9$\%	& $+0.9$\% $-0.8$\%	& $+1.6$\% $-0.20$\% \\
             $b$-tagging		& $+0.18$\% $-0.18$\%         & $+0.23$\% $-0.33$\%         & $+0.24$\% $-0.18$\%         & $<0.01$\%         & $+1.6$\% $\,<0.01$\% \\
             Luminosity				& 0.3\%	& 0.3\%	& 0.3\%	& 0.2\%	& 0.2\%\\ 
             Data/MC statistical (CR)	& 2.2\%	& 4\%	& 1.3\%	& 4\%	& 8\%\\
             \hline
             Total (stat.)              		& 2.7\%	& 5\%	& 1.5\%	& 6\%	& 10\%	\\ 
             Total (syst.)             			& 6\%	& 15\%	& 4\%	& 11\%	& 28\% \\
             \hline
             \hline
             \multicolumn{6}{c}{\textbf{Resolved}}\\
             \hline
             \textbf{Systematic uncertainty} & \textbf{SR1} & \textbf{SR2} & \textbf{SR3} & \textbf{SR4} & \textbf{SR5}\\
             \hline
             Background estimation	& 3.5\%	        & 6\%	        & 4\%	        & 8\%	        & 16\% \\ 
             $t\bar t$	cross section	& $+0.24$\% $-0.27$\%	& $+0.20$\% $-0.23$\%	& $+0.31$\% $-0.4$\%	& $+0.23$\% $-0.26$\%	& $+0.17$\% $-0.20$\% \\
             JER \smallR	& $+0.17$\%	& $+0.32$\%	& $+0.18$\%	& $-0.37$\%	& $-0.5$\% \\
             JES \smallR	& $+0.8$\% $-0.6$\%	& $+0.7$\% $-0.6$\%	& $+0.6$\% $-0.7$\%	& $+0.8$\% $-0.7$\%	& $+1.0$\% $-0.8$\%\\
             $b$-tagging	&  $+0.5$\% $-0.4$\%	& $+0.5$\% $-0.30$\%	& $+0.5$\% $-0.4$\%	& $+0.4$\% $-0.4$\%	& $+0.7$\% $-0.7$\%\\
             Luminosity           			& 0.13\%	& 0.13\%	& 0.15\%	& 0.15\%	& 0.11\%\\
             Data/MC statistical (CR)	& 1.6\%	& 2.7\%	& 1.0\%	& 3.3\%	& 6\%\\
             \hline                                
             Total (stat.)		             	& 2.1\%	& 4\%	& 1.0\%	& 4\%	& 8\%	\\
             Total (syst.)		             	& 4\%	& 7\%	& 4\%	& 8\%	& 17\%	\\
             \hline
        \end{tabular}}
        \caption{Systematic and statistical uncertainties on the total background in each of the signal regions for the merged and resolved analyses. The background estimation uncertainties have been scaled by the ratio of the multijet contribution to the total background estimation in order to get the relative error on the total background.}
        \label{tab:syst_tot_resolved_ortho}}
        \end{center}
\end{table}

%% file: Results.tex
\section{Results}
\label{sec:result}

\begin{figure}[H]
	\centering
		\subfigure[merged]{\includegraphics[width=0.47\textwidth]{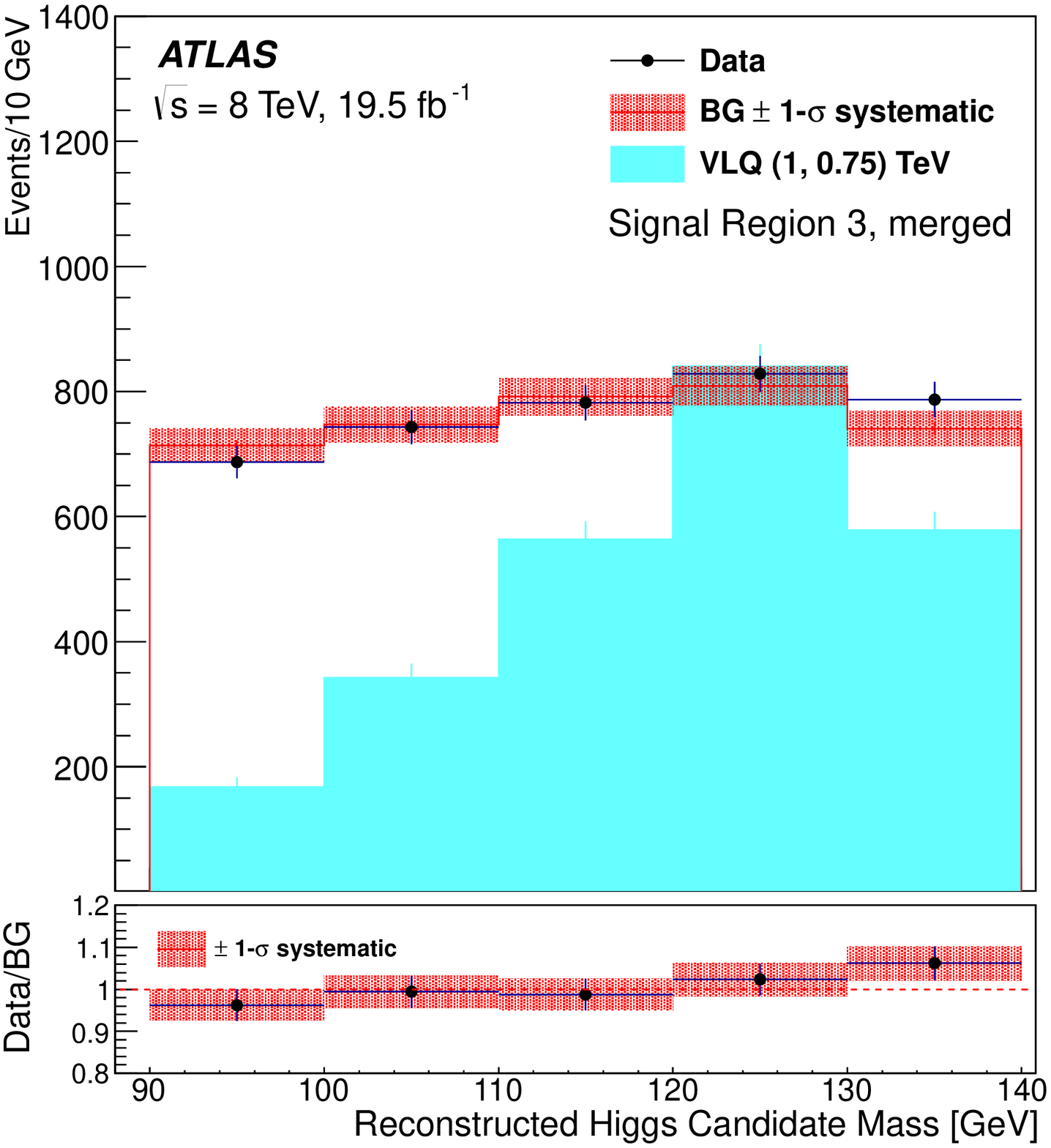}}\qquad
		\subfigure[resolved]{\includegraphics[width=0.47\textwidth]{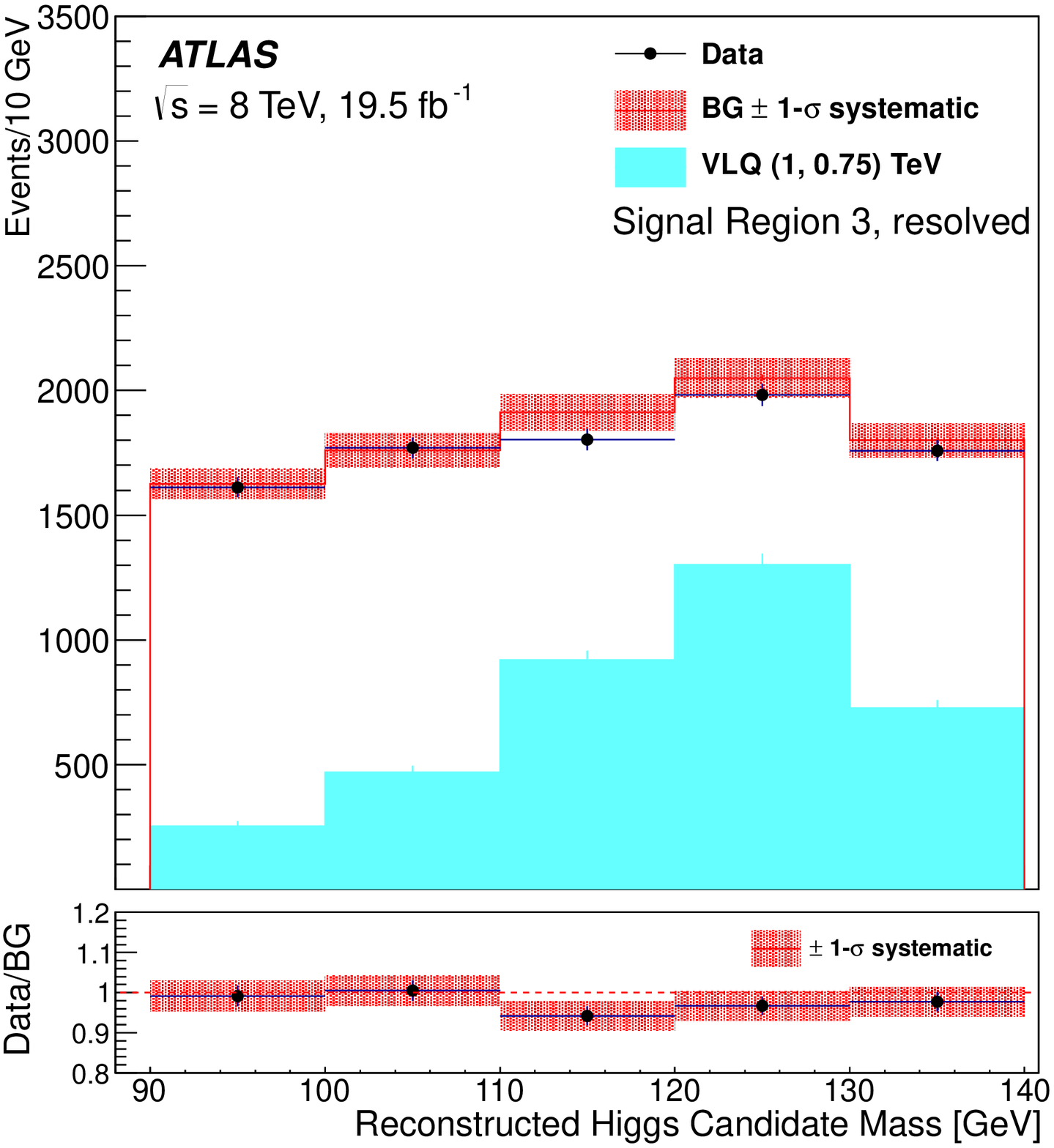}}
	\caption{Observed (black points)  and expected (red band) distribution of the reconstructed Higgs boson candidate mass in signal region~3 for the (a) merged and (b) resolved cases. The normalisation of region C is applied as an overall factor, and not bin-by-bin, for the Higgs boson candidate mass window. The red error bands represent the systematic uncertainty on the expected background. The distribution from a signal with $m_{G^*} = 1$~TeV and $m_{B_H} = 0.75$~TeV is also shown for the parameters listed in Section~\ref{sec:simulation}. The lower panels show the ratio of the observed number of events in data to the expected background.}
	\label{fig:Masses_SR3_paper}
\end{figure}

After applying all selection criteria in the signal regions,  the multijet background in the Higgs boson candidate mass window is evaluated according to Eq.~(\ref{eq:ABCD}). Mass distributions of reconstructed Higgs boson candidates are shown in Fig.~\ref{fig:Masses_SR3_paper} for the merged and resolved cases in SR3. The observed data and the background predictions are consistent within statistical and systematic uncertainties.

For each pair of mass points considered, the expected signal yield, based on the benchmark model, is evaluated in the corresponding signal region. These yields result from the signal  $\sigma \times \left(A \times \epsilon\right)$, where $\sigma$ is the cross section including all the branching fractions and $\left(A \times \epsilon\right)$ is the acceptance times reconstruction efficiency of the signal selection cuts. The amount of contamination, defined as the expected ratio of the number of signal events in control regions B, C, or D to that in the signal region, is also estimated. 

\begin{table}[H]
        \begin{center}
        \scalebox{0.90}{
        \begin{tabular}{c|c|c|c|c|c}
                \hline
                \hline
                \multicolumn{6}{c}{\textbf{Merged}}\\
                \hline
                \textbf{Background} 	& \textbf{SR1} 		& \textbf{SR2} 		& \textbf{SR3} 		& \textbf{SR4} 		& \textbf{SR5}\\
                \hline	
                Multijet	                	& $1104\pm27$	& $398\pm16$		& $3372\pm49$	& $259\pm12$		& $85\pm7$\\
                $t\bar t/\text{top}$      	& $107\pm4$		& $30.0\pm2.3$	& $398\pm8$	  	& $18.3\pm1.9$	& $4.2\pm1.0$\\
                $W/Z+\text{jets}$        	& $10.5\pm1.3$     	& $4.4\pm0.9$        	& $30.1\pm1.9$     	& $2.6\pm0.8$       	& $0.8\pm0.5$\\
                \hline
                \textbf{Total BG}    	& $1222\pm33\pm70$   & $432\pm20\pm60$      & $3800\pm60\pm150$  & $280\pm16\pm30$       & $90\pm9\pm25$\\
                \hline
                \textbf{Data}                	& 1310	           &	456      	& 3827	          & 287	                & 89\\
                \hline
                \hline
                \multicolumn{6}{c}{\textbf{Resolved}}\\
                \hline
                \textbf{Background} & \textbf{SR1} & \textbf{SR2} & \textbf{SR3} & \textbf{SR4} & \textbf{SR5}\\
                \hline
                Multijet	          	& $1985\pm34$	& $639\pm18$     	& $8580\pm90$	& $523\pm18$		& $141\pm9$\\
                $t\bar t/\text{top}$	& $64.2\pm3.2$	& $17.7\pm1.8$	& $353\pm8$		& $15.4\pm1.6$	& $3.3\pm0.7$\\
                $W/Z+\text{jets}$ 	& $35.0\pm3.3$	& $12.7\pm1.8$    	& $142\pm6$		& $12.8\pm2.2$         & $2.6\pm0.4$\\
                \hline
                \textbf{Total BG}		& $2080\pm40\pm80$	& $669\pm25\pm50$	& $9080\pm90\pm340$  & $551\pm23\pm50$       & $147\pm12\pm25$\\
                \hline
                \textbf{Data}		& 2106			& 706			& 8927			& 568			& 122\\
                \hline
        \end{tabular}}
        \caption{Observed data and background yields in the different signal regions for the merged and resolved cases. The first error is statistical and the second is systematic, while for individual background contributions only the statistical error is shown. Statistical errors on the numbers of data events in the control regions used to estimate the multijet background are included in the total statistical error. The row $t\bar t/\text{top}$ includes $t\bar t$, single-top and $t\bar t + V/H$ backgrounds while $W/Z + \text{jets}$ includes leptonic and hadronic decays of the vector boson.}
        \label{tab:mc_bg_yield}
        \end{center}
\end{table}

\begin{table}[H]
\begin{center}
\begin{tabular}{|c||c|c|c|c|c|c|c|c|c|}
\hline 
  	\boldmath$m_{B_H} \textbf{ [TeV]} \downarrow $	         & \multicolumn{9}{c|}{}		\\
 	\hline
	\hline
	&	&	&	&	&		& \cellgreen$62^{+62}_{-22}$ 	& \cellgreen$62^{+58}_{-24}$	& \cellgreen$61^{+75}_{-23}$	& \cellgreen$71^{+110}_{-29}$	\\ 
 2.0	&	&	&	&	&		& \cellgreen$68$ 			& \cellgreen$62$			& \cellgreen$64$			& \cellgreen$74$	\\ 
	&	&	&	&	&		& \cellgreen$5.2$ 			& \cellgreen$4.8$			& \cellgreen$2.9$			& \cellgreen$1.5$	\\
	\hline
		&    	&	&	&	&		&  	& 	& 	& 	\\ 
1.875	&    	&	&	&	&		&  	& 	& 	& 	\\ 
		&    	&	&	&	&		&  	& 	& 	& 	\\
	\hline
	&	&	&	&	& \cellgreen$72^{+48}_{-22}$	& \cellgreen$51^{+36}_{-17}$ 	& \cellgreen$57^{+52}_{-21}$	& \cellgreen$61^{+72}_{-24}$	& \cellgreen$65^{+82}_{-26}$	\\ 
1.75	&    	&	&	&	& \cellgreen$74$			& \cellgreen$52$ 			& \cellgreen$59$			& \cellgreen$61$				& \cellgreen$66$	\\ 
	&    	&	&	&	& \cellgreen$16$			& \cellgreen$13$ 			& \cellgreen$7.8$			& \cellgreen$3.8$			& \cellgreen$1.7$	\\
	\hline
		&    	&	&	&	&		&  	& 	& 	&	\\ 
1.625	&    	&	&	&	&		&  	& 	& 	&	\\ 
		&    	&	&	&	&		&  	& 	& 	&	\\
	\hline
	&    	&	&	&  \cellgreen$66^{+42}_{-23}$		&  \cellgreen$57^{+53}_{-19}$		& \cellgreen$57^{+55}_{-21}$ 	& \cellgreen$56^{+69}_{-21}$	& \cellgreen$56^{+47}_{-22}$	& \cellgreen$65^{+68}_{-25}$	\\ 
1.50	&    	&	&	&  \cellgreen$66$				&  \cellgreen$61$				& \cellgreen$57$ 			& \cellgreen$57$				& \cellgreen$56$			& \cellgreen$67$	\\ 
	&    	&	&	&  \cellgreen$47$				&  \cellgreen$38$				& \cellgreen$21$ 			& \cellgreen$9.6$			& \cellgreen$4.3$			& \cellgreen$1.9$	\\
	\hline
		&    	&	&	&	&	&  	& 	& \cellgreen$66^{+42}_{-23}$	& \cellyellow \\ 
 1.375	&    	&	&	&	&	&  	& 	& \cellgreen$66$			& \cellyellow \\ 
		&    	&	&	&	&	&  	& 	& \cellgreen$4.5$			& \cellyellow \\ \hline
	&    	&	&  \cellgreen$163^{+104}_{-50}$	&  \cellred$66^{+49}_{-22}$	&  \cellred$54^{+43}_{-18}$	& \cellgreen$60^{+58}_{-22}$	& \cellgreen$42^{+45}_{-15}$	& \cellyellow	& \cellyellow	\\ 
1.25	&    	&	&  \cellgreen$163$				&  \cellred$67$				&  \cellred$54$				& \cellgreen$60$		  	& \cellgreen$43$			& \cellyellow	& \cellyellow	\\ 
	&    	&	&  \cellgreen$148$				&  \cellred$105$			&  \cellred$54$				& \cellgreen$24$		  	& \cellgreen$11$			& \cellyellow	& \cellyellow	\\
	\hline
		&    	&	&	&	&	&  \cellgreen$68^{+63}_{-23}$	& \cellyellow	& \cellyellow	& \cellyellow	\\ 
1.125	&    	&	&	&	&	&  \cellgreen$70$			& \cellyellow	& \cellyellow	& \cellyellow	\\ 
		&    	&	&	&	&	&  \cellgreen$25$			& \cellyellow	& \cellyellow	& \cellyellow	\\
	\hline
	&    	&  \cellred$157^{+76}_{-52}$	&  \cellred$151^{+87}_{-50}$	&  \cellred$53^{+37}_{-18}$	&  \cellred$58^{+46}_{-19}$	& \cellyellow	& \cellyellow	& \cellyellow	& \cellyellow	\\ 
1.0	&    	&  \cellred$152$			&  \cellred$151$			&  \cellred$54$				&  \cellred$58$			& \cellyellow	& \cellyellow	& \cellyellow	& \cellyellow	\\ 
	&    	&  \cellred$475$			&  \cellred$291$			&  \cellred$159$			&  \cellred$58$			& \cellyellow	 & \cellyellow	& \cellyellow	& \cellyellow	\\
	\hline
		&    	&	&	& \cellred$84^{+44}_{-26}$	& \cellyellow	& \cellyellow	& \cellyellow	& \cellyellow	& \cellyellow	\\ 
0.875	&    	&	&	& \cellred$85$				& \cellyellow	& \cellyellow	& \cellyellow	& \cellyellow	& \cellyellow	\\ 
		&    	&	&	& \cellred$137$			& \cellyellow	& \cellyellow	& \cellyellow	& \cellyellow	& \cellyellow	\\
	\hline
	&\cellred$232^{+130}_{-73}$	& \cellred$172^{+90}_{-52}$ 	& \cellred$59^{+31}_{-19}$	& \cellyellow	& \cellyellow	& \cellyellow	& \cellyellow	& \cellyellow	& \cellyellow	\\ 
0.75	&\cellred$232$    			& \cellred$172$ 			& \cellred$59$ 				& \cellyellow	& \cellyellow	& \cellyellow	& \cellyellow	& \cellyellow	& \cellyellow	\\ 
	&\cellred$1449$  			& \cellred$746$ 			& \cellred$310$ 			& \cellyellow	& \cellyellow	& \cellyellow	& \cellyellow	& \cellyellow	& \cellyellow	\\
	\hline
	\hline
 	\boldmath$m_{G^*} \textbf{ [TeV]} \to          $         &  1.0	 	& 1.25	& 1.5 	&	1.75 &	2.0 &  2.25	& 2.50	& 2.75	& 3.0 \\
	\hline
\end{tabular}
\end{center}
\caption{Combined limits, in fb, on $\sigma\left(pp \to G^* \to B_H b\right) \times \text{BR}\left(B_H \to Hb\right) \times \text{BR}\left(H \to b \bar b\right)$. First and second entries in each cell give the expected and observed limits, respectively. The third entry gives the cross section in fb predicted by the benchmark model. Red cells are excluded and green cells are not excluded at 95\%~C.L. Cases where $m_{G^*} > 2m_{B_H}$ are not considered in this analysis and are marked in yellow.}
\label{tab:2Dlimits_combined}
\end{table}

\begin{figure}[H]                                                                                                                                                       
        \begin{center}
                \includegraphics[width=0.8\textwidth]{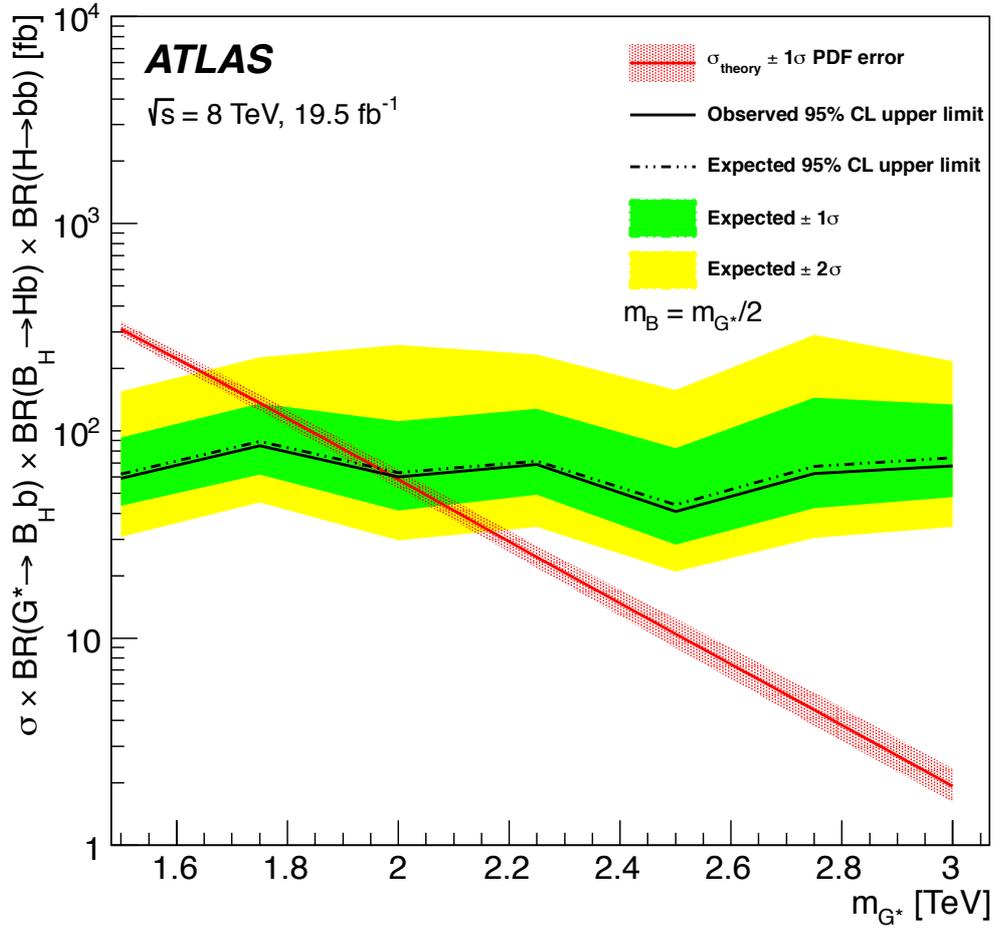}
                \end{center}
        \caption{Observed (solid) and expected (dashed) 95\% C.L. upper limits on the cross section $\sigma\left(pp \to G^* \to B_H b\right) \times \text{BR}\left(B_H \to Hb\right) \times \text{BR}\left(H \to b \bar b\right)$ for VLQ mass points with $m_{B_H} = m_{G^*}/2$, from the combined merged and resolved analyses, as well as the theoretical prediction based on parameters given in Section~\ref{sec:simulation}. The uncertainty band around the theory cross section reflects the uncertainty in the CTEQ6.6 PDFs.}
        \label{fig:limitCombined}
\end{figure}

\begin{figure}[H]                                                                                                                                                         
        \begin{center}
                \includegraphics[width=0.8\textwidth]{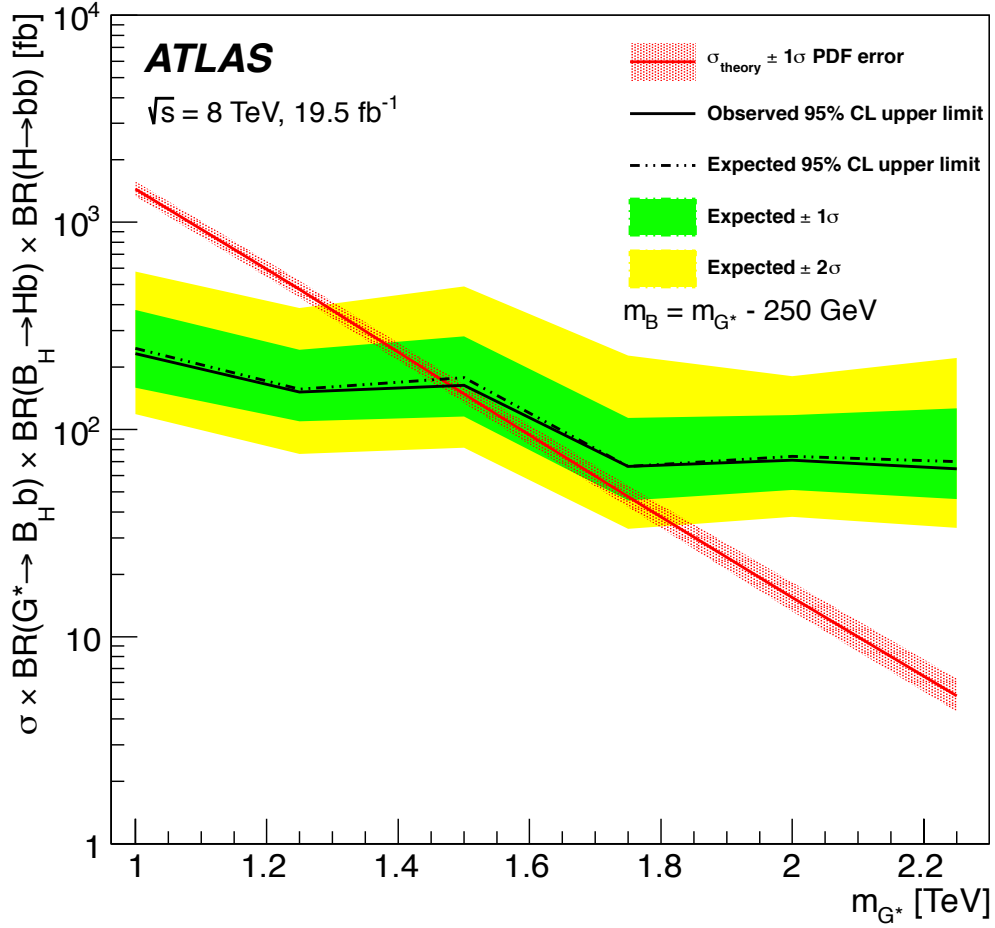}
                \end{center}
        \caption{Observed (solid) and expected (dashed) 95\% C.L. upper limits on the cross section $\sigma\left(pp \to G^* \to B_H b\right) \times \text{BR}\left(B_H \to Hb\right) \times \text{BR}\left(H \to b \bar b\right)$ for VLQ mass points with $m_{B_H} = m_{G^*} - 250$~GeV, from the combined merged and resolved analyses, as well as the theoretical prediction based on parameters given in Section~\ref{sec:simulation}. The uncertainty band around the theory cross section reflects the uncertainty in the CTEQ6.6 PDFs.}
        \label{fig:limitCombined2}
\end{figure}

Table~\ref{tab:mc_bg_yield} shows the expected and observed background event yields in each of the signal regions for the merged and resolved cases. No significant excess of data events is found compared to the expected SM background. Taking into account the number of expected background events in each of the signal regions and the yield of signal events for each test mass pair, together with all statistical and systematic uncertainties, upper limits at the 95\% confidence level (CL), using the $\text{CL}_{\rm S}$ prescription~\cite{Read:2002hq} and RooStats~\cite{Schott:2012zb}, are set on the cross section times the branching fraction of a signal, combining results from the merged and resolved analyses. To account for possible contamination of the control regions by signal, an iterative procedure is used:  a 95\% CL limit is first obtained assuming no contamination in the control regions. The contamination in regions B, C, D is then calculated, assuming a signal corresponding  to that limit, and the multijet background is then re-evaluated. The procedure is repeated until it converges to a stable value. Expected and observed limits on the cross section $\sigma(pp \to G^* \to B_Hb) \times \text{BR}(B_H \to Hb) \times \text{BR}(H \to b \bar b)$, where $\sigma(pp \to G^* \to B_Hb)$ represents the cross section of the process $pp \to G^* \to B_H\bar b$ and its complex conjugate, as well as the theoretical cross section for the benchmark model, with its theoretical uncertainty, are shown in Table~\ref{tab:2Dlimits_combined}. Limits for the particular cases where $m_{B_H} = m_{G^*}/2$ and $m_{B_H} = m_{G^*} - 250$~GeV are shown in Figs.~\ref{fig:limitCombined} and~\ref{fig:limitCombined2}.

%% file: Conclusion.tex
\section{Conclusion}
\label{sec:conclusion}

A search for a heavy gluon and a charge $-1/3$ vector-like quark in the process $pp \to G^*\to B_H \bar b$, with $B_H \to b H$ and $H \to b\bar b$, has been performed using an integrated luminosity of $19.5\text{ fb}^{-1}$ of $pp$ collision data recorded at $\sqrt s = 8$~TeV with the ATLAS detector at the LHC. The main background, multijet production, is estimated with a data-driven technique. Five signal regions are defined based on the choice of jet assignment to the $B_H$ candidate and on lower mass requirements for the reconstructed \Glu and $B_H$. No significant excess over the SM predictions is observed and upper limits have been set at the 95\% confidence level on the total cross section times branching ratio in the two-dimensional plane of $m_{G^*}$ $\text{vs.}$ $m_{B_H}$ with $m_{G^*} \le 2m_{B_H}$. Using a benchmark model presented in Ref.~\cite{Chala:2013ega}, a lower limit of 2.0~TeV on the $G^*$ mass is obtained when $m_{G^*} = 2m_{B_H}$.

%% file: acknowledgements/Acknowledgements.tex

We thank CERN for the very successful operation of the LHC, as well as the
support staff from our institutions without whom ATLAS could not be
operated efficiently.

We acknowledge the support of ANPCyT, Argentina; YerPhI, Armenia; ARC, Australia; BMWFW and FWF, Austria; ANAS, Azerbaijan; SSTC, Belarus; CNPq and FAPESP, Brazil; NSERC, NRC and CFI, Canada; CERN; CONICYT, Chile; CAS, MOST and NSFC, China; COLCIENCIAS, Colombia; MSMT CR, MPO CR and VSC CR, Czech Republic; DNRF and DNSRC, Denmark; IN2P3-CNRS, CEA-DSM/IRFU, France; GNSF, Georgia; BMBF, HGF, and MPG, Germany; GSRT, Greece; RGC, Hong Kong SAR, China; ISF, I-CORE and Benoziyo Center, Israel; INFN, Italy; MEXT and JSPS, Japan; CNRST, Morocco; FOM and NWO, Netherlands; RCN, Norway; MNiSW and NCN, Poland; FCT, Portugal; MNE/IFA, Romania; MES of Russia and NRC KI, Russian Federation; JINR; MESTD, Serbia; MSSR, Slovakia; ARRS and MIZ\v{S}, Slovenia; DST/NRF, South Africa; MINECO, Spain; SRC and Wallenberg Foundation, Sweden; SERI, SNSF and Cantons of Bern and Geneva, Switzerland; MOST, Taiwan; TAEK, Turkey; STFC, United Kingdom; DOE and NSF, United States of America. In addition, individual groups and members have received support from BCKDF, the Canada Council, CANARIE, CRC, Compute Canada, FQRNT, and the Ontario Innovation Trust, Canada; EPLANET, ERC, FP7, Horizon 2020 and Marie Sk{\l}odowska-Curie Actions, European Union; Investissements d'Avenir Labex and Idex, ANR, R{\'e}gion Auvergne and Fondation Partager le Savoir, France; DFG and AvH Foundation, Germany; Herakleitos, Thales and Aristeia programmes co-financed by EU-ESF and the Greek NSRF; BSF, GIF and Minerva, Israel; BRF, Norway; the Royal Society and Leverhulme Trust, United Kingdom.

The crucial computing support from all WLCG partners is acknowledged
gratefully, in particular from CERN and the ATLAS Tier-1 facilities at
TRIUMF (Canada), NDGF (Denmark, Norway, Sweden), CC-IN2P3 (France),
KIT/GridKA (Germany), INFN-CNAF (Italy), NL-T1 (Netherlands), PIC (Spain),
ASGC (Taiwan), RAL (UK) and BNL (USA) and in the Tier-2 facilities
worldwide.

%% file: atlas_authlist.tex
\begin{flushleft}
{\Large The ATLAS Collaboration}

\bigskip

G.~Aad$^\textrm{\scriptsize 85}$,
B.~Abbott$^\textrm{\scriptsize 112}$,
J.~Abdallah$^\textrm{\scriptsize 150}$,
O.~Abdinov$^\textrm{\scriptsize 11}$,
B.~Abeloos$^\textrm{\scriptsize 116}$,
R.~Aben$^\textrm{\scriptsize 106}$,
M.~Abolins$^\textrm{\scriptsize 90}$,
O.S.~AbouZeid$^\textrm{\scriptsize 136}$,
H.~Abramowicz$^\textrm{\scriptsize 152}$,
H.~Abreu$^\textrm{\scriptsize 151}$
% ...
\bigskip
\\
% ...
$^{11}$ Institute of Physics, Azerbaijan Academy of Sciences, Baku, Azerbaijan\\
% ...
$^{85}$ CPPM, Aix-Marseille Universit{\'e} and CNRS/IN2P3, Marseille, France\\
% ...
$^{90}$ Department of Physics and Astronomy, Michigan State University, East Lansing MI, United States of America\\
% ...
$^{106}$ Nikhef National Institute for Subatomic Physics and University of Amsterdam, Amsterdam, Netherlands\\
% ...
$^{112}$ Homer L. Dodge Department of Physics and Astronomy, University of Oklahoma, Norman OK, United States of America\\
% ...
$^{116}$ LAL, Univ. Paris-Sud, CNRS/IN2P3, Universit{\'e} Paris-Saclay, Orsay, France\\
% ...
$^{136}$ Santa Cruz Institute for Particle Physics, University of California Santa Cruz, Santa Cruz CA, United States of America\\
% ...
$^{150}$ Institute of Physics, Academia Sinica, Taipei, Taiwan\\
$^{151}$ Department of Physics, Technion: Israel Institute of Technology, Haifa, Israel\\
$^{152}$ Raymond and Beverly Sackler School of Physics and Astronomy, Tel Aviv University, Tel Aviv, Israel\\
% ...
\end{flushleft}
